\documentclass[runningheads]{llncs}
\usepackage{subcaption}

\usepackage{xcolor}
\usepackage{soul}
\usepackage{hyperref}

\usepackage{graphicx}
\usepackage{colortbl}
\usepackage{hhline}
\usepackage{float}
\usepackage{mathptmx}  

\begin{document}
%
\title{Eyes on teleporting: comparing locomotion techniques in Virtual Reality with respect to presence, sickness and spatial orientation}
\titlerunning{Eyes on teleporting}
%
\author{Ariel Caputo\inst{1}\orcidID{0000-0002-6478-4663} \and
Massimo Zancanaro\inst{2,3}\orcidID{0000-0002-1554-5703} \and
Andrea Giachetti\inst{1}\orcidID{0000-0002-7523-6806}}
\authorrunning{A. Caputo et al.}
%
\institute{University of Verona, Department of Engineering for Innovation Medicine
\and
University of Trento, Department of Psychology and Cognitive Science
\and
Fondazione Bruno Kessler, FBK, Trento
}
\maketitle              

\begin{center}\fbox{\parbox[c]{0.7\textwidth}{
    {\centering\scriptsize preprint version - please cite as\\

Caputo, A., Zancanaro, M., Giachetti, A. (2023). Eyes on Teleporting: Comparing Locomotion Techniques in Virtual Reality with Respect to Presence, Sickness and Spatial Orientation. 

In Abdelnour Nocera, J., Kristín Lárusdóttir, M., Petrie, H., Piccinno, A., Winckler, M. (eds) Human-Computer Interaction INTERACT 2023. Lecture Notes in Computer Science, vol 14144. Springer. https://doi.org/10.1007/978-3-031-42286-7\_31
\\
Published version: https://link.springer.com/chapter/10.1007/978-3-031-42286-7\_31
    }}}    
\end{center}

\begin{abstract}
This work compares three locomotion techniques for an immersive VR environment: two different types of teleporting (with and without animation) and a manual (joystick-based) technique. We tested the effect of these techniques on visual motion sickness, spatial awareness, presence, subjective pleasantness, and perceived difficulty of operating the navigation. We collected eye tracking and head and body orientation data to investigate the relationships between motion, vection, and sickness. Our study confirms some results already discussed in the literature regarding the reduced invasiveness and the high usability of instant teleport while increasing the evidence against the hypothesis of reduced spatial awareness induced by this technique. We reinforce the evidence about the issues of extending teleporting with animation. Furthermore, we offer some new evidence of a benefit to the user experience of the manual technique and the correlation of the sickness felt in this condition with head movements.  The findings of this study contribute to the ongoing debate on the development of guidelines on navigation interfaces in specific VR environments.
\end{abstract}

\section{Introduction}

As immersive virtual reality applications are becoming widespread, a better understanding of the effects of different locomotion techniques on user experience is paramount. Although there is an increasing amount of research on this topic (see below for a short review, and Butussi \& Chittaro \cite{buttussi_locomotion_2021} and Al Zayer and colleagues \cite{al2018virtual} for recent literature analysis), there is still a stringent need to collect and compare more studies in different contexts of usages and along different dimensions of user experience. 

The term (virtual) locomotion technique refers to a control technique for allowing a person to move freely in a virtual environment while remaining within a small physical space \cite{templeman1999virtual}. It is well known that virtual movements in an immersive environment can induce sickness \cite{chang2020virtual} and the specific locomotion technique can have an impact on that \cite{dong2011control,frommel2017effects,buttussi_locomotion_2021}, together with other individual variables such as individual propensity to sickness \cite{golding_motion_1998,stanney1999motion}, and sex \cite{munafo_virtual_2017}. Furthermore, the choice of locomotion technique can impact the experience of presence, although the evidence for that is less clear \cite{buttussi_locomotion_2021,frommel2017effects,bowman1997travel,bozgeyikli2016point}.

Several locomotion techniques have been proposed in the literature, but manual movement and teleporting are the two most often investigated in research and employed in consumer applications \cite{al2018virtual,buttussi_locomotion_2021}. The manual locomotion technique involves using a joystick or a similar controller to move the viewport in the chosen direction and control the speed by operating the controller. Teleporting is more straightforward, requiring the user to select a destination point in the environment and confirm it to move the viewpoint to the target. 


Teleporting can be implemented in two ways: 
instant teleporting that can reduce the sickness (as the visual motion perceived is removed) at the cost of losing spatial awareness or teleporting with animation (where the final position is reached by moving the viewport, typically at a constant speed on a straight path). 
Previous studies tried to assess both the advantages of teleporting to reduce sickness with respect to manual control \cite{bozgeyikli2016point}, to compare manual control and teleporting with respect to both sickness and orientation \cite{bowman1997travel,bhandari2018teleportation} or to evaluate the effects on orientation and sickness of continuous vs. discrete animation when using teleporting \cite{rahimi2018scene,adhikari2022integrating}. 

However, none of these studies directly compare the manual control with both instant and animated teleporting in an active exploration task, which might be relevant since manual control (i.e. the use of a joystick to move the viewport) is the most simple and straightforward locomotion technique. In Rahimi et al. \cite{rahimi2018scene} the discrete animation is evaluated in a task featuring a passive viewport change, in \cite{adhikari2022integrating} jumps are added to continuous viewpoint control and not to teleport.

Furthermore, even if most studies feature subjective and objective measurements quantifying sickness and orientation, they do not evaluate the relationships of these effects with head and eye movements, which can be relevant to understand the causes of the sickness and suggesting strategies to reduce it.

In this paper, we designed a user study that directly compare all the three  locomotion approaches for navigation and orientation tasks also evaluating the behavior of the subjects in terms of eyes, head and body movements, and considering the experienced sickness and the susceptibility to sickness of the participants.



%

Specifically, we aimed to investigate the following research questions:

$Q1$: is there a relation between sickness and susceptibility to sickness, and do the specific locomotion methods affect this relation differently?

$Q2$: do the different locomotion techniques impact efficiency, spatial awareness, the perception of presence, difficulty, and pleasantness of the task? 

$Q3$: do the behaviour (in terms of eyes and body movements) change among the different conditions, and do those changes explain presence, difficulty, or pleasantness?

The study involved 25 volunteer participants who had the task to navigate a virtual environment resembling a museum with each of the three locomotion techniques. The study was approved by the Person Research Approval Committee (CARP) of the University of Verona.

\section{Related works}

There are many studies in the literature regarding the perception of discomfort while navigating in virtual reality and many techniques have been proposed to reduce this effect. Recent surveys \cite{tian2022review,chang2020virtual}, however, pointed out that many findings of these studies are inconsistent. Chang and colleague \cite{chang2020virtual} suggest that multiple factors of a VR system are related to users’ discomfort, while Tian et al. \cite{tian2022review} pointed out that there is a huge variability in the sickness effects in different users and it is necessary to profile individual susceptibility, and there is lack of proper validated physiological measurements.

The control of the navigation is an important factor affecting the perceived sickness. Several methods have been proposed to provide efficient ways of navigating VR without inducing VR sickness \cite{al2018virtual}.  
A method to avoid it is certainly the use of teleporting, as also shown in the work of Frommel et al \cite{frommel2017effects} and Bozgeyikli et al.\cite{bozgeyikli2016point}. Teleporting is considered a valid solution for in-place VR locomotion, as it is also more efficient and usable than other methods \cite{buttussi_locomotion_2021}. However, locomotion techniques that instantly teleport users to new locations may increase user disorientation \cite{bowman1997travel}.

If a user wants to navigate in an exploratory way while avoiding sickness and maintaining spatial orientation, it is not obvious whether to use teleportation or a method of reducing sickness by controlling navigation more freely.

A few recent studies address the simultaneous evaluation of sickness and spatial orientation for different sets of locomotion control methods and tasks, even if with some limitations.
Langbehn et al. \cite{langbehn2018evaluation} compared joystick based control, similar to our manual technique, with instant teleport and redirected walking, but only for limited movements on pre-defined paths. They found that redirected walking provided a better understanding of the space and that joystick-based navigation did not provide better orientation while resulting in higher sickness.

In a work of Bhandari et al. \cite{bhandari2018teleportation}, teleportation with a fast animation at a constant speed, similar to our teleporting with animation, is proposed as a possible solution to avoid disorientation. 
While the authors adopt this solution to limit the exposure to optical flow and found it effective, it must be noted, looking at the environment used for the tests, that the virtual environment used for the test presents limited texturing and no obstacles, so that the optical flow is not too relevant and this may have biased the results.

Rahimi and colleagues \cite{rahimi2018scene} 
compared three variations of teleporting (instant, animated and pulsed) evaluating effects in spatial awareness, in an object-tracking task involving three moving objects with a viewport transition not controlled by the user.
They concluded that animation might improve spatial awareness, but it also worsens sickness. The results on the passive task are not necessarily the same that would have obtained in an active one. 

Farmani and Theater \cite{farmani2020evaluating} tested the effects of rotation and translation snapping, showing that discontinuous motion could result in reduced sickness with no disorientation effects. The effects of introducing the discontinuity, however, was studied separately for the different motion on simple tasks avoiding body motions.

Adhikari et al. \cite{adhikari2022integrating} proposed to add iterative jumps (i.e. ‘HyperJump’) to continuous locomotion reaching faster navigation in large environments but without significant changes in orientation and sickness. The discontinuous animation was not, however, applied to teleport and/or compared with simple teleport. A similar idea was also proposed by Riecke \cite{riecke2022hyperjumping}, who merged continuous locomotion with jumps in a novel navigation control interface.

While all these studies provide useful ideas and insights on how to reduce sickness and mantain spatial awareness, we have seen that they presents some limitations making often not obvious to derive guidelines for the navigation control design in classic exploration tasks in VR applications.
Furthermore, many studies are based on the assumption that the methods to reduce the sickness should be based on the  

reduction of the visual illusion of self-motion often referred to as \emph{vection} \cite{palmisano2015future}, there is evidence of a decoupling of vection and motion sickness \cite{webb2003eye,kuiper2019vection}. 
\color{black}
According to Webb et al., \cite{webb2003eye}, vection is influenced by the peripheral vision, while visual-induced motion sickness seems to be influenced by central (foveal) visual stimulation or by eye movements, in particular, by the optokinetic response (e.g., the automatic movement of the eye following objects in motion when the head remains stationary) \cite{webb2002optokinetic}. The relationships between sickness and vection are complex, and the outcomes of related studies are not always considered in the literature on methods to reduce motion sickness in immersive  Virtual Reality.

Keshavarz et al., \cite{keshavarz2015vection}, tackled these issues and pointed out that "vection might be a necessary prerequisite for visual-induced motion sickness, but only in combination with other factors (e.g., sensory conflict, postural stability, eye movements, head movements, etc."

This suggest that user studies aimed at assessing motion sickness in conjunction with other factors (e.g. orientation) should capture as much as possible fine-grained physiological measurements to understand the origin of the different symptoms.
In general, there is a need for further studies to understand the complex relationships between locomotion control and user experience, as well as sickness, vection, and eye and body movements. This fact motivated us to design a novel experiment to investigate on them.

\begin{figure} [t]
    \centering
    \begin{subfigure}[b]{0.42\textwidth}
    \includegraphics[width=1\linewidth]{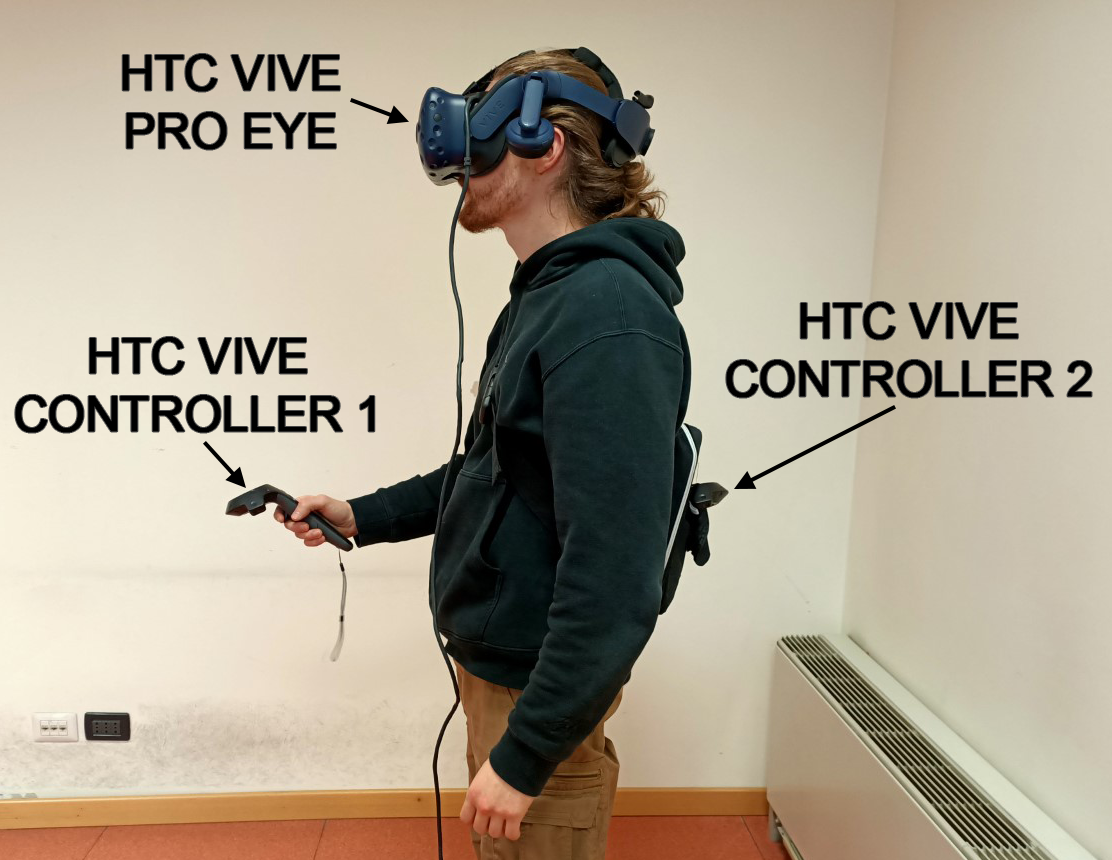}
    \caption{ }\label{fig:setup}
     \end{subfigure}
     \hspace{1em}
      \begin{subfigure}[b]{0.48\textwidth}
    \includegraphics[width=1\linewidth]{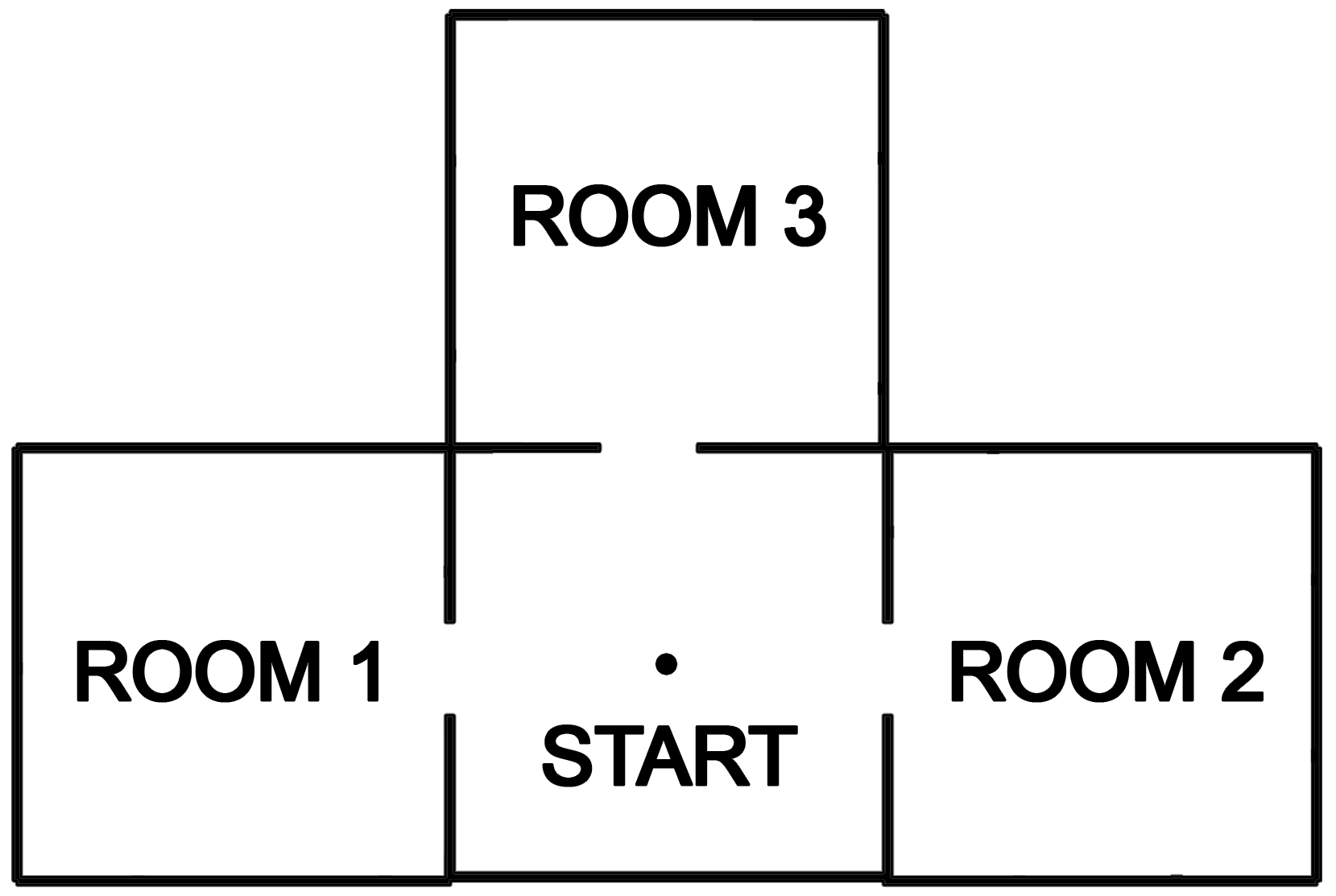}
    \caption{ }
   \label{fig:museumlayout}
       \end{subfigure}
        \label{}
         \caption{(a) Hardware setup (HMD and Controller 1 for interaction and Controller 2 to track the torso orientation), (b) Map of the museum used in the orientation task} 
\end{figure}

\section{The study}

In order to assess the research questions outlined above, we designed a task to be performed in an immersive VR environment, and we asked the participants to experience it with each of the three locomotion techniques. The study had a within-subject design with 25 participants and it was exploratory in nature. 

\subsection{Materials and task}
\begin{figure*}[ht]

      \centering
    \begin{subfigure}[b]{0.46\textwidth}
    \includegraphics[width=\textwidth]{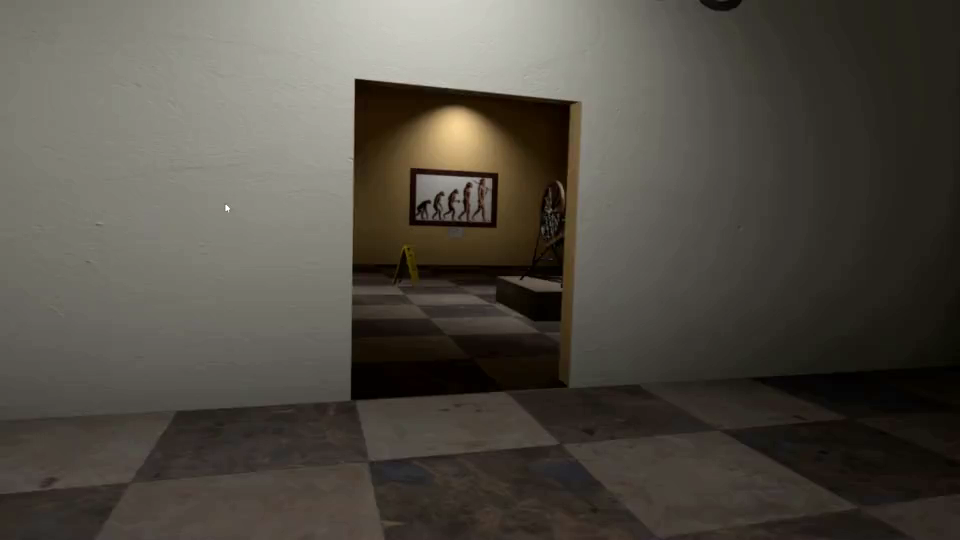}
    \caption{}\label{fig:red}
    \end{subfigure}
    \begin{subfigure}[b]{0.46\textwidth}
    \includegraphics[width=\textwidth]{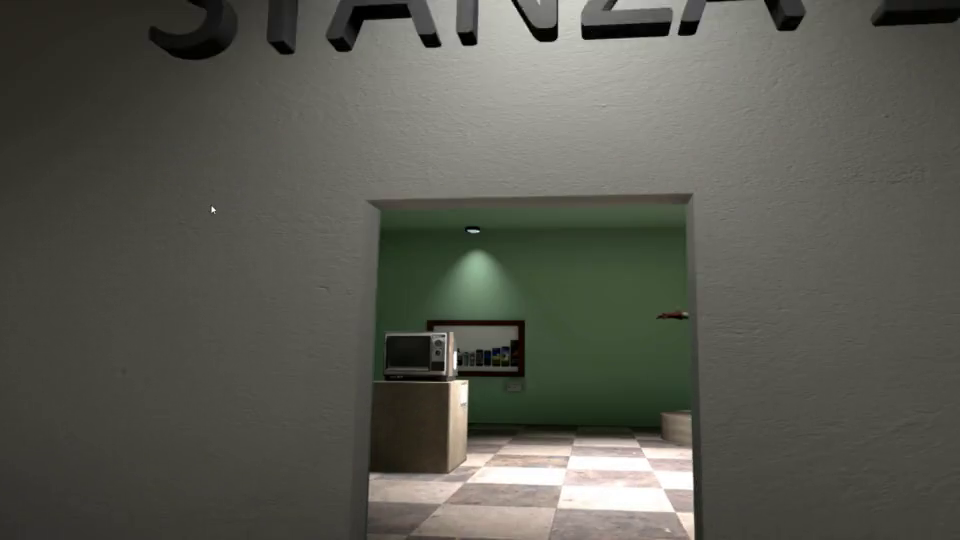}
    \caption{}\label{fig:b}
    \end{subfigure}
    \begin{subfigure}[b]{0.46\textwidth}
    \includegraphics[width=\textwidth]{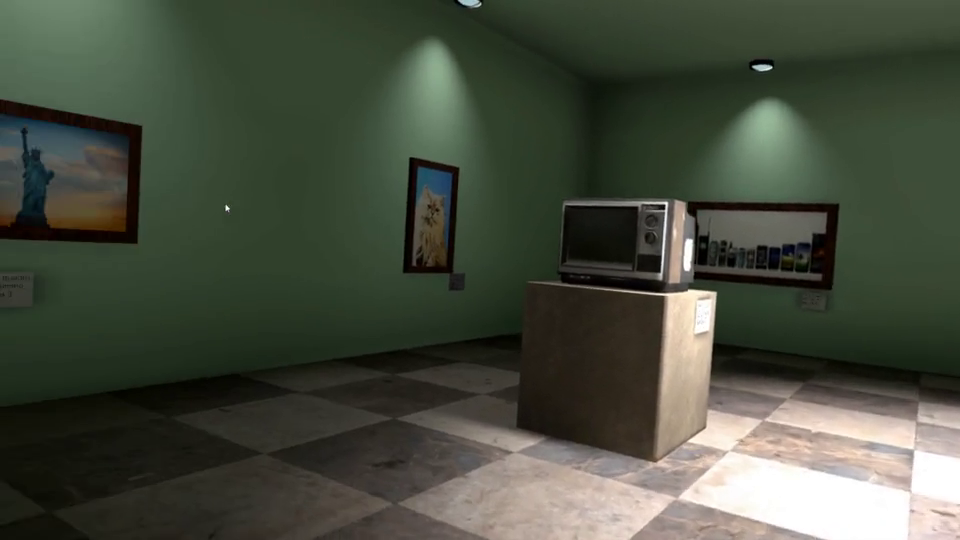}
    \caption{}\label{fig:c}
    \end{subfigure}
       \begin{subfigure}[b]{0.46\textwidth}
    \includegraphics[width=\textwidth]{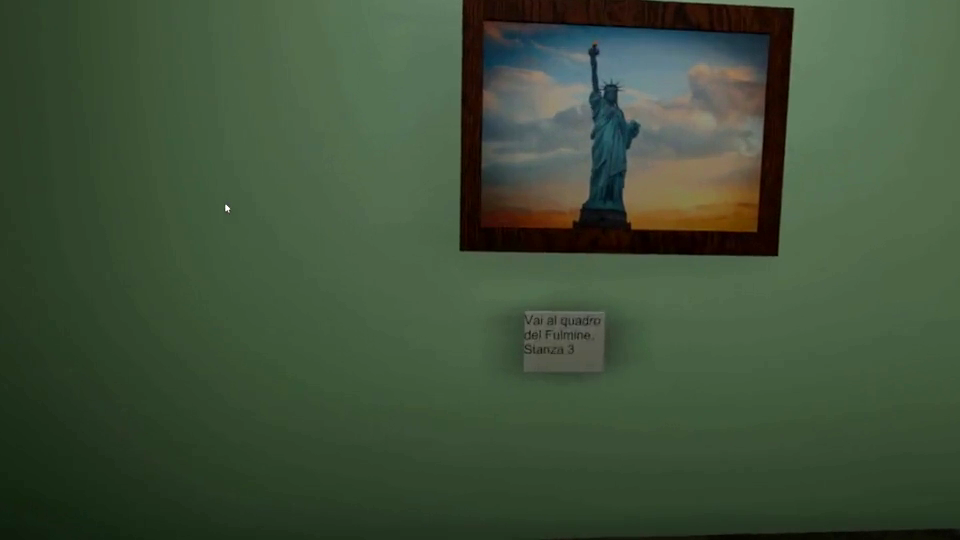}
    \caption{}\label{fig:plate}
    \end{subfigure}
    \begin{subfigure}[b]{0.46\textwidth}
    \includegraphics[width=\textwidth]{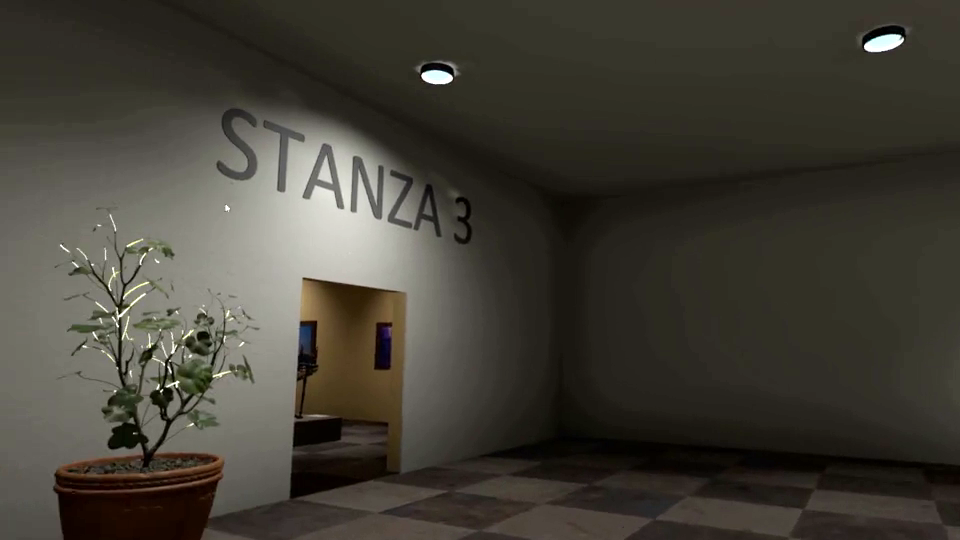}
    \caption{}\label{fig:prop}
    \end{subfigure}
    \begin{subfigure}[b]{0.46\textwidth}
    \includegraphics[width=\textwidth]{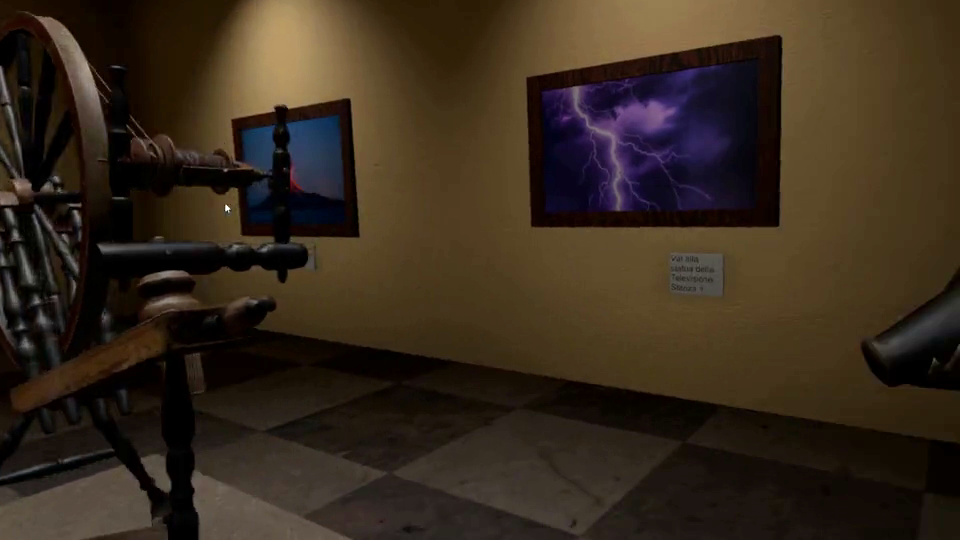}
    \caption{}\label{fig:plate2}
    \end{subfigure}
    \caption{Navigation task in the virtual museum. (a) Starting position in front of room 3.  The subject is asked to go in room 2 (b,c) find the painting with the Statue of Liberty (d) and read the plaque under it. The text asks to go in room 3 (e), find the painting with the lightning  and follow the instructions there reported, iterating the procedure (f).}
    \label{fig:navi}
\end{figure*}

We used the HTC Vive Pro Eye Head-Mounted Display for our experiments, as it allows the recording of gaze data and head movements. Subjects used an HTC Vive controller to guide the locomotion, while a second controller was positioned on the back of the user to automatically record the torso orientation data (\autoref{fig:setup}). 
The application was developed using the Unity framework and featured a Virtual Museum environment with 4 rooms arranged as shown in \autoref{fig:museumlayout}. 
Each room contains 6 paintings distributed along the walls and 3 items, such as sculptures or other props, displaced around the room to create non-trivial navigation options (\autoref{fig:navi}). Each item is associated with a plaque attached close to it on the wall or on the pedestal.

We used these plaques to define the navigation task. In fact, the plaque associated with each item displayed the instructions to reach the next one to be visited (for example, "Go visit painting with the lighting in Room 3" (\autoref{fig:plate2}). These instructions are arranged such that the user visits all the items in the museum returning to the starting item and creating a looping path.

The task required the subjects to navigate for a fixed amount of time (5 minutes), and we counted the number of reached items to assess the locomotion efficiency.

To avoid order effects across different conditions for the same participant, we created three loops. Each loop starts from a different initial item, which is communicated to the user when the task begins.


As explained above, the application implemented 3 different locomotion techniques (which were the three conditions of the study):

\begin{itemize}
    \item \textbf{Manual:}  This technique implements a straightforward mapping of the movements captured by the Vive Controller's directional pad. On the vertical axis, we added the forward and backward directions of the movement while, on the horizontal axis, the left and right strafing can be used. The speed of movement, regardless of the direction, is 7 km/h. The forward direction is aligned with the camera direction, meaning the reference frame of the movement controls is the same as the camera's, projected on a plane parallel to the museum floor. In other words, the technique recreates the mechanics of the character's movements used in many popular video games featuring a first-person view. \color{black}
    \item \textbf{Teleport:}
    our implementation of the instant Teleport locomotion control uses the tracked Vive Controller: when the user presses and holds the trigger button, a visible ray, directed along the device axis, appears to help the user to identify the target position (\autoref{fig:plate}). While the user can aim the device at any visible part of the environment, the viewport translation only takes place upon releasing the trigger button, provided that the ray points towards an empty spot on the floor. In order to provide feedback about the potential outcome of the teleport command, the ray is displayed as green if the target spot is valid (\autoref{fig:statue}) or red if not (\autoref{fig:red}). If the button is released when the pointed target is valid, the viewport is then instantly translated to the x-y coordinates of the target.

    \begin{figure}
    \centering
     \centering
    \begin{subfigure}[b]{0.49\textwidth}
    \includegraphics[width=\textwidth]{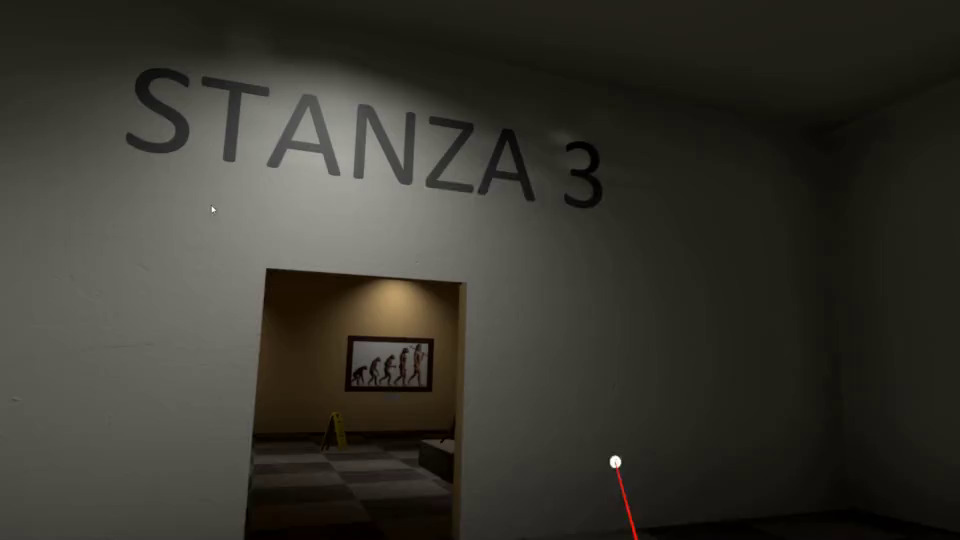}
    \caption{}\label{fig:start}
    \end{subfigure}
    \begin{subfigure}[b]{0.49\textwidth}
    \includegraphics[width=\textwidth]{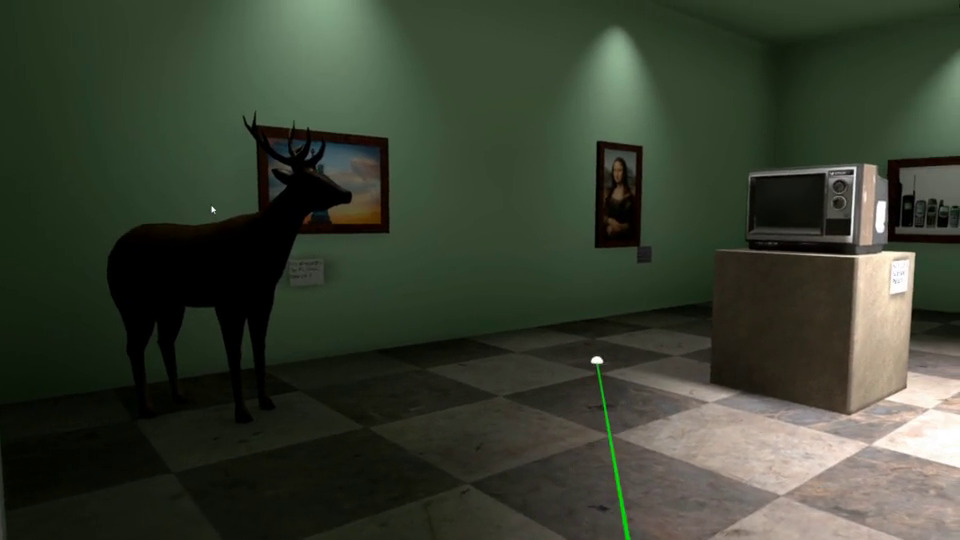}
    \caption{}\label{fig:statue}
    \end{subfigure}
    \caption{(a) Teleport ray aimed at a not reachable position and colored in red. (b) When the ray is aimed at a reachable position the ray is colored in green and when the button is released the viewpoint is moved to the target.}
     \label{fig:tele}
\end{figure}

    \item \textbf{Teleport+animation:} This technique is a variation of the Teleport technique described above. The destination selection and visual feedback (the color of the ray) are the same. The difference is that, after the release of the trigger button, the viewport position is not instantly changed, but rather it is moved following a straight path towards the selected spot at a constant speed of 7 km/h. This value has been chosen because it is a realistic human walking speed, although slightly higher than the average \cite{fitzpatrick2006another}. Preliminary pilot tests confirmed that it resulted comfortable for users. In fact, we tested higher speeds like those proposed in \cite{bhandari2018teleportation}, but they were not comfortable for users in our museum scenario. Both in the preliminary tests and the user study several participants reported slight discomfort in using the animated teleport because of the feeling of potential collisions with walls and statues. 
    
\end{itemize}

The main task assigned to the participants consisted of navigating the museum for a fixed amount of time (5 minutes) and visiting the different items. Each subject performed the task three times, each time with a different locomotion technique and following a different path. 
The paths started with the first instruction given by the experiment supervisor, asking to reach a particular item in a room, read the associated plate, go to the artwork indicated in the text, read its plate, and so on. 
To remove the biases due to the execution order with the different paths and locomotion techniques, the technique-path pair for each task was assigned following a Latin square arrangement. In this way, the same number of users had a particular path or technique in the first, second, and third trials.

We also assigned a secondary task, asking the subjects, at the end of each run, to indicate the position of 2 paintings and one sculpture on a map representing the museum's layout (shown in \autoref{fig:museumlayout}). 
The paintings were chosen from the items the user had to visit, while the statue was just placed in one of the rooms and never mentioned in the instructions. We introduced this task to assess the effects of the different techniques on spatial orientation and awareness and eventually evaluate possible trade-offs with other effects, such as motion sickness.

\subsection{Participants and measures}
We collected data with questionnaires (pre- and post-study) and automatically collected data from the VR application runs. 
In detail, the app recorded the head orientation, the torso orientation, and the eye movements captured by the Vive tracking systems. The app also counted the number of visited items for each navigation task. 

We used Motion Sickness Susceptibility Questionnaire (MSSQ) for the pre-study measure of susceptibility to sickness, the scores were normalized in percentiles \cite{golding_motion_1998}.

After each task performed in the different conditions, we administered a questionnaire to the subjects to assess the experienced sickness, the perceived presence, and task difficulty. 
The questions include those of the standard  Simulator Sickness Questionnaire (SSQ) from which we estimated the scales for nausea, oculomotor, and disorientation \cite{kennedy1993simulator}. While we are aware that the use of SSQ to measure visual-induced motion sickness has been recently criticized \cite{hirzle2021critical}, for this study, we decided to keep using it for the sake of comparison with the extant literature on the topic.

To evaluate the perceived presence, we used the questions of the igroup presence questionnaire (IPQ), and measured the scales of spatial presence, involvement, and realness, adding a single item about the sense of being in a place \cite{schubert_experience_2001}.

After the conclusion of all the tasks, we asked the subjects to fill out a final questionnaire, to have an overall assessment of the perceived difficulty and pleasantness on a 5-point Likert scale.

\begin{table}
\centering
\resizebox{\columnwidth}{!}{
\begin{tabular}{|l|c|c|c|c|c|c|} 
\hhline{~------|}
\multicolumn{1}{l|}{}                                         & \multicolumn{2}{c|}{{\cellcolor[rgb]{0.8,0.8,0.8}}\textbf{Manual }}            & \multicolumn{2}{c|}{{\cellcolor[rgb]{0.8,0.8,0.8}}\textbf{Teleport }}          & \multicolumn{2}{c|}{{\cellcolor[rgb]{0.8,0.8,0.8}}\textbf{Teleport+anim }}      \\ 
\hhline{~------|}
\multicolumn{1}{l|}{}                                         & {\cellcolor[rgb]{0.675,0.78,0.894}}Mean & {\cellcolor[rgb]{0.89,0.89,0.89}}STD & {\cellcolor[rgb]{0.675,0.78,0.894}}Mean & {\cellcolor[rgb]{0.89,0.89,0.89}}STD & {\cellcolor[rgb]{0.675,0.78,0.894}}Mean & {\cellcolor[rgb]{0.89,0.89,0.89}}STD  \\ 
\hline
{\cellcolor[rgb]{0.8,0.8,0.8}}\textbf{SSQ Nausea}             & {\cellcolor[rgb]{0.902,0.949,1}}1.85    & 0.67                                 & {\cellcolor[rgb]{0.902,0.949,1}}1.37    & 0.57                                 & {\cellcolor[rgb]{0.902,0.949,1}}2.04    & 0.85                                  \\ 
\hline
{\cellcolor[rgb]{0.8,0.8,0.8}}\textbf{SSQ Oculomotor}         & {\cellcolor[rgb]{0.902,0.949,1}}1.70    & 0.51                                 & {\cellcolor[rgb]{0.902,0.949,1}}1.35    & 0.54                                 & {\cellcolor[rgb]{0.902,0.949,1}}1.80    & 0.53                                  \\ 
\hline
{\cellcolor[rgb]{0.8,0.8,0.8}}\textbf{SSQ Disorientation}     & {\cellcolor[rgb]{0.902,0.949,1}}2.77    & 0.85                                 & {\cellcolor[rgb]{0.902,0.949,1}}2.02    & 0.79                                 & {\cellcolor[rgb]{0.902,0.949,1}}2.79    & 0.81                                  \\ 
\hline
{\cellcolor[rgb]{0.8,0.8,0.8}}\textbf{IPQ Spatial Presence}   & {\cellcolor[rgb]{0.902,0.949,1}}4.18    & 0.85                                 & {\cellcolor[rgb]{0.902,0.949,1}}4.12    & 1.02                                 & {\cellcolor[rgb]{0.902,0.949,1}}3.88    & 0.87                                  \\ 
\hline
{\cellcolor[rgb]{0.8,0.8,0.8}}\textbf{IPQ Involvement}        & {\cellcolor[rgb]{0.902,0.949,1}}3.58    & 1.01                                 & {\cellcolor[rgb]{0.902,0.949,1}}3.89    & 0.90                                 & {\cellcolor[rgb]{0.902,0.949,1}}3.71    & 0.89                                  \\ 
\hline
{\cellcolor[rgb]{0.8,0.8,0.8}}\textbf{IPQ Realness}           & {\cellcolor[rgb]{0.902,0.949,1}}2.74    & 0.64                                 & {\cellcolor[rgb]{0.902,0.949,1}}2.48    & 0.68                                 & {\cellcolor[rgb]{0.902,0.949,1}}2.44    & 0.78                                  \\ 
\hline
{\cellcolor[rgb]{0.8,0.8,0.8}}\textbf{Post Difficulty}           & {\cellcolor[rgb]{0.902,0.949,1}}2.08    & 0.83                                 & {\cellcolor[rgb]{0.902,0.949,1}}1.46    & 1.02                                 & {\cellcolor[rgb]{0.902,0.949,1}}2.71    & 1.12                                  \\ 
\hline
{\cellcolor[rgb]{0.8,0.8,0.8}}\textbf{Post Pleasantness}           & {\cellcolor[rgb]{0.902,0.949,1}}3.67    & 1.09                                 & {\cellcolor[rgb]{0.902,0.949,1}}4.17    & 0.96                                 & {\cellcolor[rgb]{0.902,0.949,1}}1.96    & 1.00                                  \\ 
\hline
{\cellcolor[rgb]{0.8,0.8,0.8}}\textbf{Objects Visited}        & {\cellcolor[rgb]{0.902,0.949,1}}19.04   & 8.58                                 & {\cellcolor[rgb]{0.902,0.949,1}}20.71   & 8.64                                 & {\cellcolor[rgb]{0.902,0.949,1}}16.25   & 6.85                                  \\ 
\hline
{\cellcolor[rgb]{0.8,0.8,0.8}}\textbf{Objects Located}        & {\cellcolor[rgb]{0.902,0.949,1}}0.83   & 0.92                                 & {\cellcolor[rgb]{0.902,0.949,1}}0.75   & 0.74                                 & {\cellcolor[rgb]{0.902,0.949,1}}1.13   & 1.15                                  \\ 
\hline
{\cellcolor[rgb]{0.8,0.8,0.8}}\textbf{Cumulative Torso Angle} & {\cellcolor[rgb]{0.902,0.949,1}}103.45  & 32.13                                & {\cellcolor[rgb]{0.902,0.949,1}}132.03  & 86.53                                & {\cellcolor[rgb]{0.902,0.949,1}}119.83  & 42.02                                 \\ 
\hline
{\cellcolor[rgb]{0.8,0.8,0.8}}\textbf{Cumulative Head Angle}  & {\cellcolor[rgb]{0.902,0.949,1}}173.15  & 39.76                                & {\cellcolor[rgb]{0.902,0.949,1}}298.54  & 63.54                                & {\cellcolor[rgb]{0.902,0.949,1}}216.04  & 41.28                                 \\ 
\hline
{\cellcolor[rgb]{0.8,0.8,0.8}}\textbf{Total Eye Movement}     & {\cellcolor[rgb]{0.902,0.949,1}}4643.82 & 1780.43                              & {\cellcolor[rgb]{0.902,0.949,1}}6264.48 & 1796.92                              & {\cellcolor[rgb]{0.902,0.949,1}}6126.62 & 1719.97                               \\
\hline
\end{tabular}
}
\vskip\baselineskip
\caption{Means and standard deviations for different measurements per navigation methods. Angles are reported in radians and eye movements were measured on the screen space plane with range [-1,1] on both axis.}
\label{tab:means}
\end{table}

\section{Results}
Of the 25 participants, one has been excluded because of some errors in data collection. The remaining 24 subjects (14 males and 10 females) completed the required tasks. Most had no or few previous experiences with VR (respectively 9, and 10), while 4 participants declared some previous experience, and the remaining 1) had several previous experiences. Being recruited on the university campus, the large majority of the participants (21) were in the 20-35 years old age range.

In the following, we  present the descriptive statistics for the collected data, including reliability analysis for the multi-items questionnaires and  normality check for all the variables. The Mann–Whitney test (equivalent of a non-parametric t test) has been used to check the male vs female possible differences. The inferential analysis related to the research questions is presented in subsection 4.1. 
\color{black}

We also controlled the possible effect of sex on the other measures because the extant literature has contradictory results (\cite{munafo_virtual_2017}. Although our study did not aim at investigating this difference, it was necessary to check  its possible impact on the other measures. 
\color{black}

The susceptibility to sickness collected with the MMSQ questionnaire was low ($mean=26.02$, $std=27.91$) and not normally distributed (Shapiro $s=0.81$, $p<0.01$). There was no difference between male and female participants (Mann–Whitney $w=80$, $p=0.576$).

 Looking at the sickness experienced by the participants, all the three scales used from the SSQ questionnaire had good internal reliability (specifically, $0.803$ for the nausea scale, $0.815$) for the oculomotor scale and $0.66$ for the disorientation scale. All the scales were not normally distributed (Shapiro, $s=0.875$ , $p<0.01$ for nausea; $s=0.907$ , $p<0.01$ for oculomotor; and $s=0.901$ , $p<0.01$ for disorientation). There were no differences between males and females in either of the three scales (Mann–Whitney $w=673$, $p=0.621$ for nausea; Mann–Whitney $w=702$, $p=0.409$ for oculomotor; and, Mann–Whitney $w=719$, $p=0.303$ for disorientation). 

For the IPQ questionnaire on presence, the reliability of the spatial presence scale was low (Cohen’s $alpha=0.154$) mainly because of one item that was consequently eliminated, the adjusted scale reached then the satisfactory alpha of $0.708$. The reliability of the involvement scale was low too (Cohen’s $alpha=0.386$) and also in this case the elimination of one single item raised the reliability to a satisfactory alpha of $0.633$. The reliability of the realness scale was low too but, in this case, all the items had low correlation among themselves: we, therefore, decided not to use this sub-scale.
The scales of spatial presence and involvement were not normally distributed (Shapiro $s=0.969$, $p=0.077$ for spatial presence; $s=0.972$, $p=0.106$ for involvement). There were no statistical differences between males and females (Mann–Whitney $w=644$, $p=0.877$ for spatial presence; $w=580.5$, $p=0.574$ for involvement).

As objective measures, we first counted the number of artworks visited during each task and the number of objects correctly located in the post-task questionnaire.
The count of the artworks visited in 5 minutes was normally distributed (Shapiro, $s=0.924$, $p<0.01$),  with no statistical differences between males and females (Mann–Whitney $w=728.0$, $p=0.265$).
as well as the correctly located query item count (Shapiro, $s=0.801$, $p<0.01$) with no statistical differences between males and females (Mann–Whitney $w=564.0$, $p=0.423$).

\begin{figure}[bt]
    \centering
  \begin{subfigure}{0.45\textwidth}
    \centering
   \includegraphics[width=\textwidth]{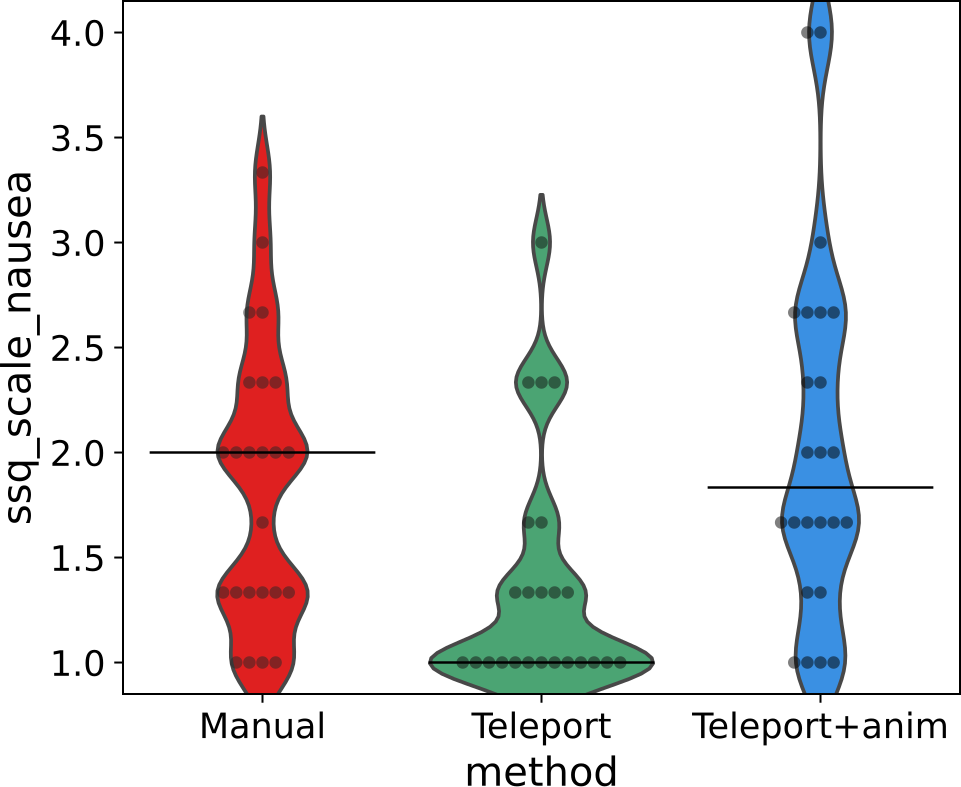}
     \caption{}\label{fig:hr}
    \end{subfigure}
    \vskip 10pt
    \begin{subfigure}{0.45\textwidth}
    \centering
      \includegraphics[width=\textwidth]{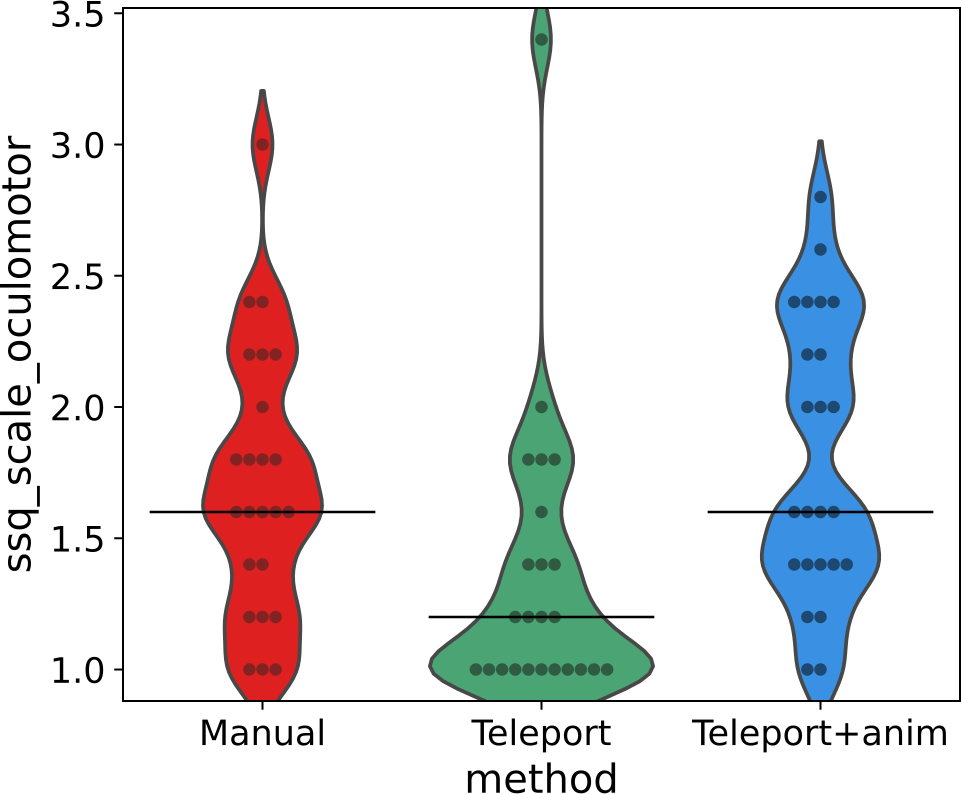}
    \caption{}\label{fig:tr}
    \end{subfigure}
      \begin{subfigure}{0.45\textwidth}
    \centering
      \includegraphics[width=\textwidth]{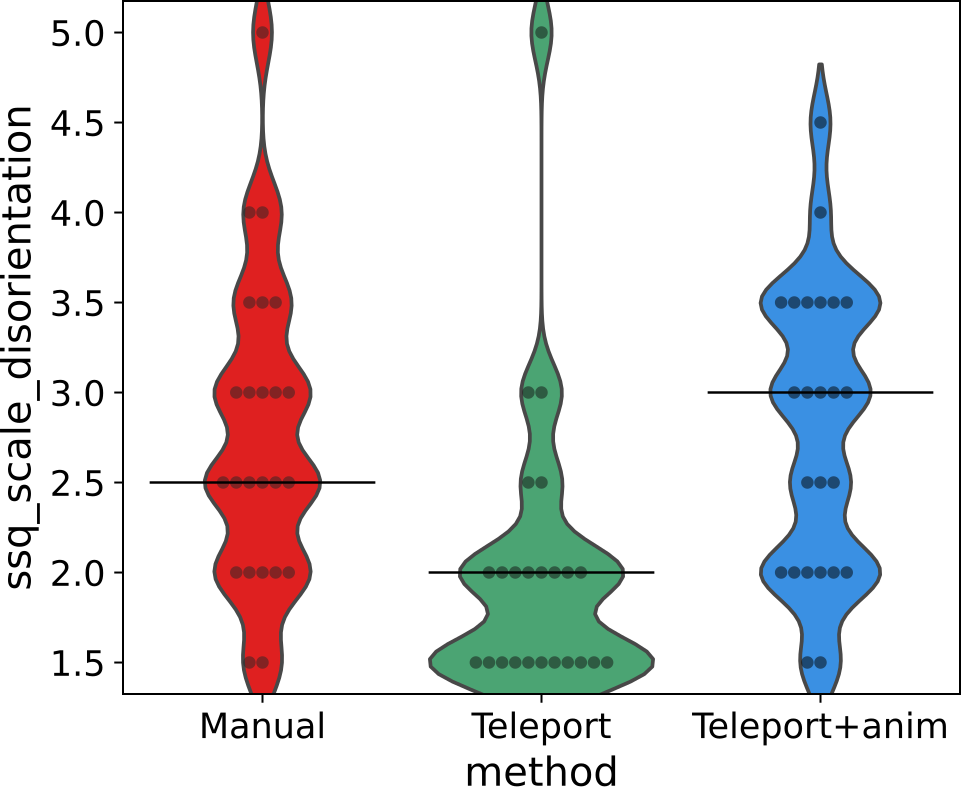}
    \caption{}\label{fig:tr}
    \end{subfigure}
 \caption{Violin plot of the scales used to measure the effect of the sickness (nausea, oculomotor and disorientation) grouped by navigation method.}
    \label{fig:sickness}
\end{figure}


Concerning the eye, head, and body tracking data, we estimated the accumulated amount of eye movements and the orientation changes for both the torso and the head during the tasks in the three experimental conditions. For the torso, the measures of 6 subjects were missed in some conditions, therefore in the analysis below, the remaining 18 participants only are used for what concerns torso angle variations. The distribution of the eye movements was not normally distributed (Shapiro $s=0.971$, $p=0.094$) while both the distribution of the angles of torso and head were normally distributed (Shapiro $s=0.95$, $p<0.05$ for angle variations of the torso; and, $s=0.952$, $p<0.01$ for the angle variations of the head). Yet, in both cases, there is a statistically significant difference in the variances (Bartlett’s K-squared, $k=17.415$, $p<0.01$ for the torso and $k=6.554$, $p<0.05$ for the head), therefore we decided to use non-parametric statistics for these measures too. There were no statistical differences between males and females (Mann–Whitney $w=467$, $p=0.063$ for eye movements; $w=480.5$, $p=0.915$ for torso angle; and, $w=537$, $p=0.291$ for head angle).

 Finally, for what concerns the post-study measures of perceived difficulty and perceived pleasantness, they were both not normally distributed (Shapiro $s=0.837$, $p<0.01$ for difficulty; and, $s=0.881$, $p<0.01$ for pleasantness). No statistical differences between males and females (Mann–Whitney $w=733$, $p=0.219$ for difficulty; and, $w=651$, $p=0.810$ for pleasantness).

The mean scores and standard deviations for all the measures are reported in \textbf{\autoref{tab:means}}.


\subsection{Inferential statistics}

For the inferential statistics, we mainly used Friedman test as a non-paramtric equivalent of the repeated measures ANOVA, since the data are usually not normally distributed. To assess correlations, we first employed linear mixed models and then, when significant, we used regression analysis to estimate the direction.  As good practice in statistical analysis, we always reported non significant outcomes too.  
\color{black}

With respect to $Q1$, all the three measures of sickness are different in the three conditions (Friedman $chi=16.692$ $p<0.01$ for nausea; $chi=23.718$, $p<0.01$ for disorientation; and, $chi=17.452$, $p<0.01$ for oculomotor). Pairwise comparisons using Conover’s all-pairs test with Bonferroni correction reveal that the teleporting technique scores significantly lower with respect to the other two techniques, while there is no difference between manual and teleport+animation (\autoref{fig:sickness}).

\begin{figure}[h]
    \centering
    \includegraphics[width=0.9\linewidth]{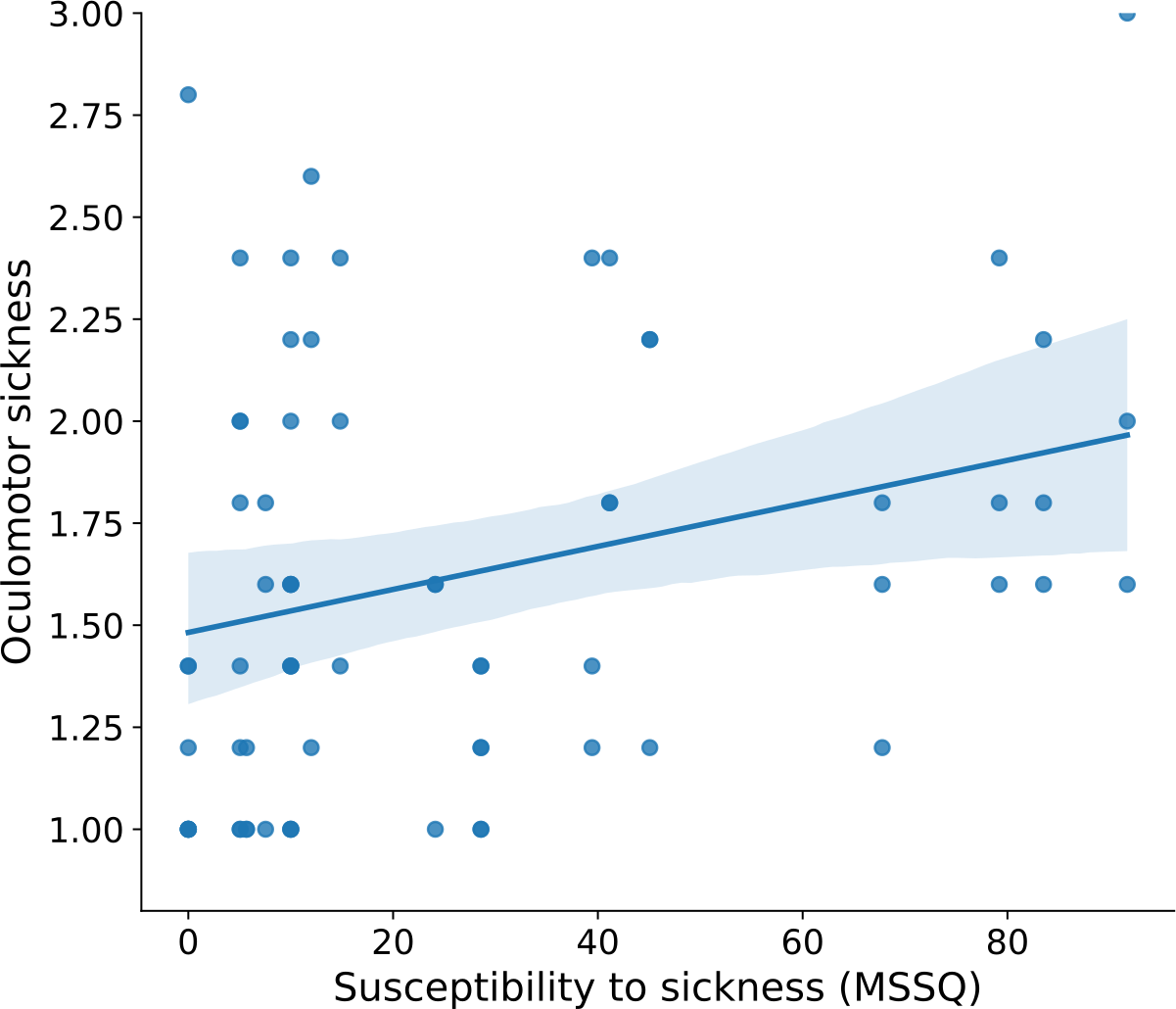}
    \caption{Regression model for the oculomotor sickness (SSQ) and susceptibility to sickness (MSSQ).}
    \label{fig:ssq_mssq}
\end{figure}

We found that, among the correlations between each of the sickness measures and the susceptibility to sickness, the only one statistically different from zero is the oculomotor scale for the manual technique (Pearson’s product-moment correlation $t=2.144$, $p<0.05$). Yet, a regression model between these variables demonstrates that the explained variance is relatively low (adjusted R squared $0.135$) (\autoref{fig:ssq_mssq}).

Considering the research question $Q2$, for what concerns presence, there are no statistical differences on either the scale of spatial presence, or involvement (respectively, Friedman $chi=5.448$, $p=0.066$ for spatial presence; and, $chi=0.026$, $p=0.987$ for involvement). Yet, for the general presence (the sense of being in place \cite{heeter_being_1992}), there is a small but significant difference (Friedman $chi=12.299$, $p<0.01$). Pairwise comparisons using Conover’s all-pairs test with Bonferroni correction show that teleport+animation is lower than manual ($p<0.01$) but the other comparisons are not significant.
\begin{figure}[th]
    \centering
\includegraphics[width=0.65\linewidth]{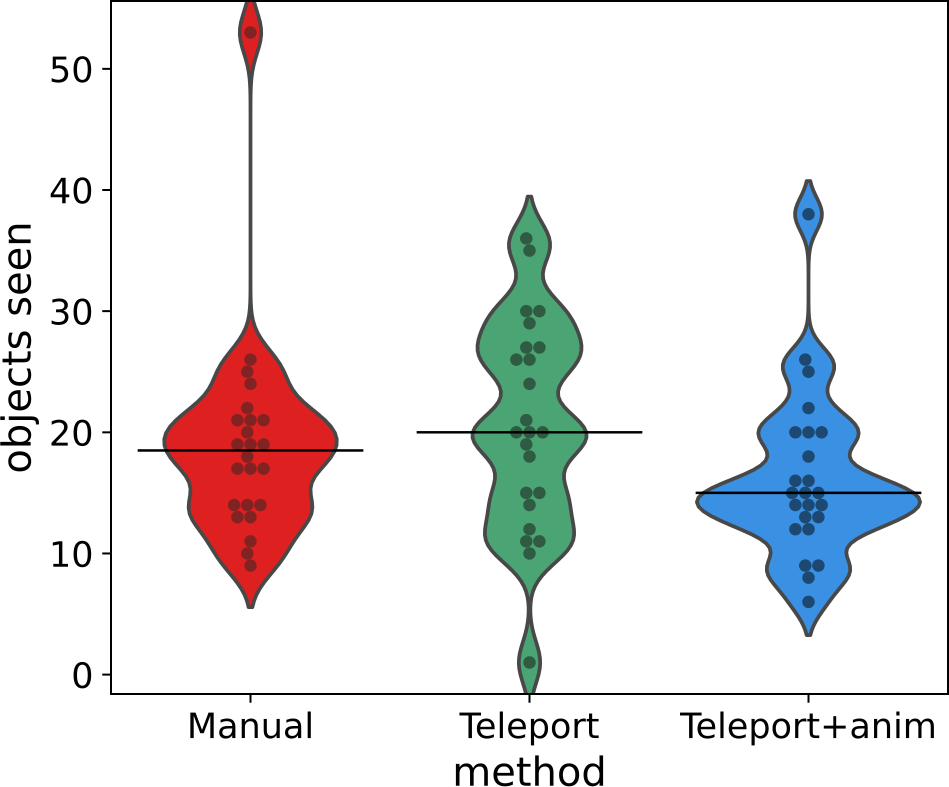}
     \caption{Violin plot of the number of objects visited by the users grouped by navigation method.}\label{fig:seen}
\end{figure}
In the perceived difficulty scores, there is a difference among the three conditions (Friedman $chi=16.892$, $p<0.01$). Pairwise comparisons using Conover’s all-pairs test with Bonferroni correction show that teleport is perceived as less difficult than teleport+animation but the other comparisons do not reach significance.

For the perceived pleasantness, there is a difference among the three conditions (Friedman $chi=28.787$, $p<0.01$). Pairwise comparisons using Conover’s all-pairs test with Bonferroni correction show that both teleport and manual are perceived as more pleasant than teleport+animation and there is no statistical difference between manual and teleport.

\begin{figure}[!h]
    \centering
      \includegraphics[width=0.68\linewidth]{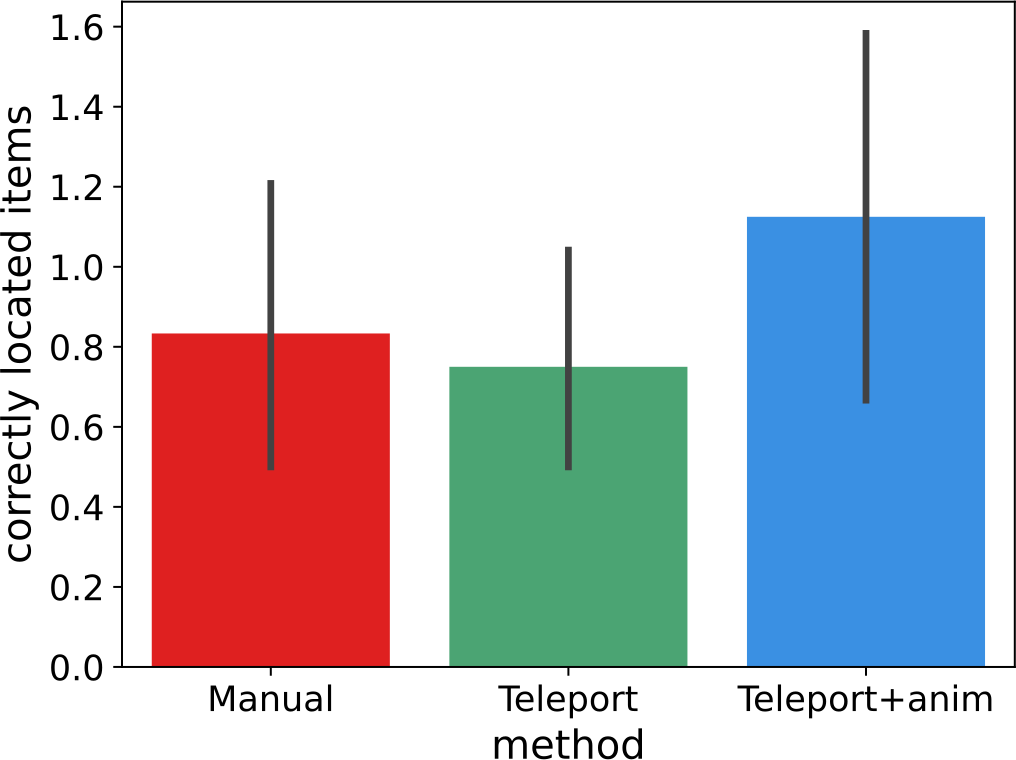}
    \caption{Average number of objects correctly located on the map by the subjects at the end of each run, grouped by navigation method.}\label{fig:located}
\end{figure}

\begin{figure}[!h]
    \centering
    \begin{subfigure}[b]{0.6\linewidth}
    \centering
      \includegraphics[width=\linewidth]{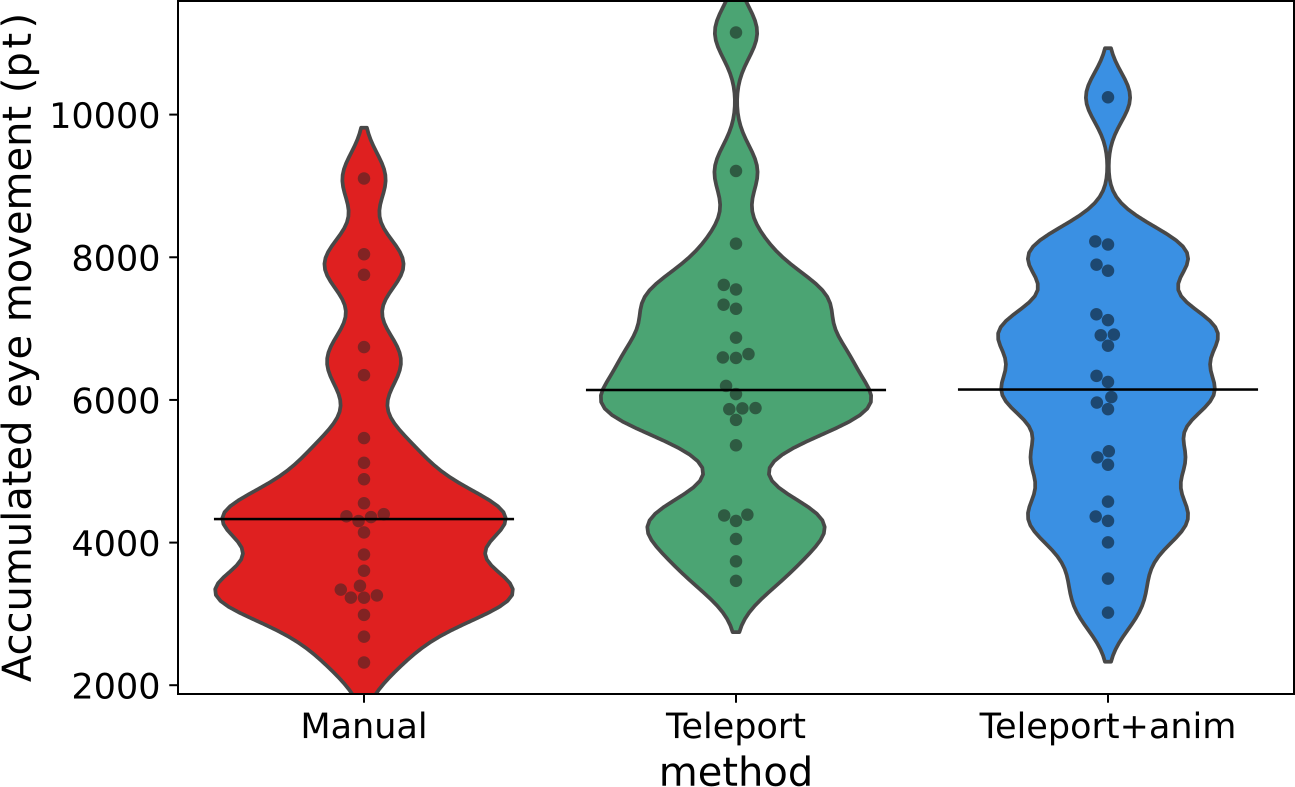}
    \caption{}\label{fig:eyem}
    \end{subfigure}
    \vskip 10pt
  \begin{subfigure}[b]{0.42\linewidth}
    \centering
   \includegraphics[width=\linewidth]{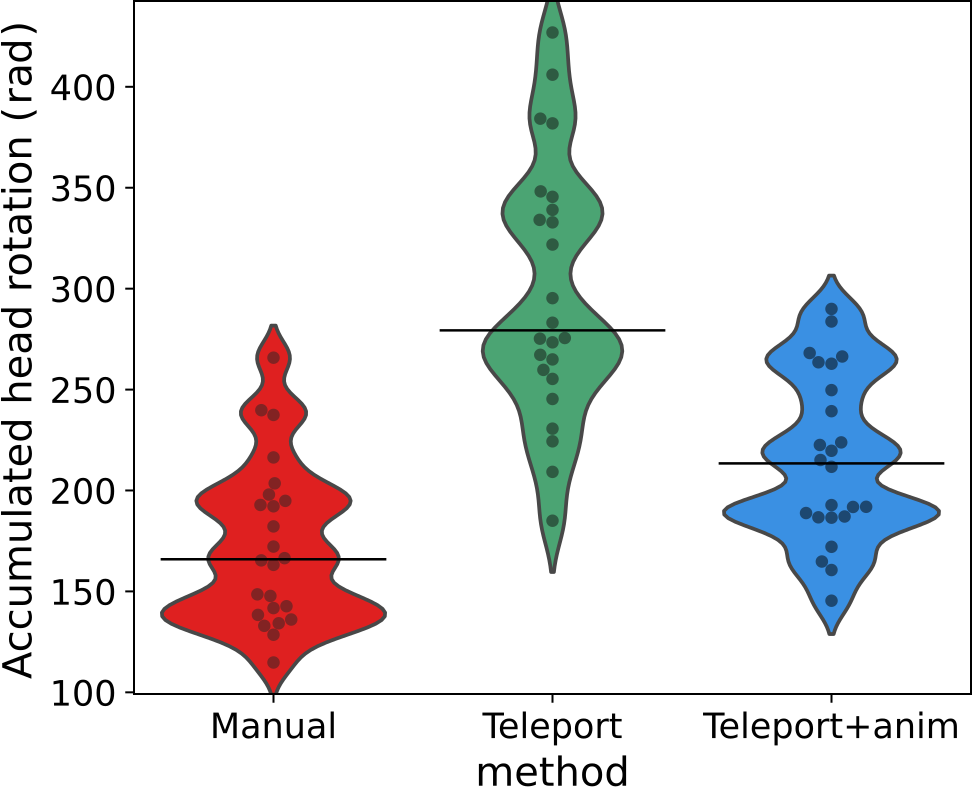}
     \caption{}\label{fig:headr}
    \end{subfigure}
      \begin{subfigure}[b]{0.54\linewidth}
    \centering
      \includegraphics[width=\linewidth]{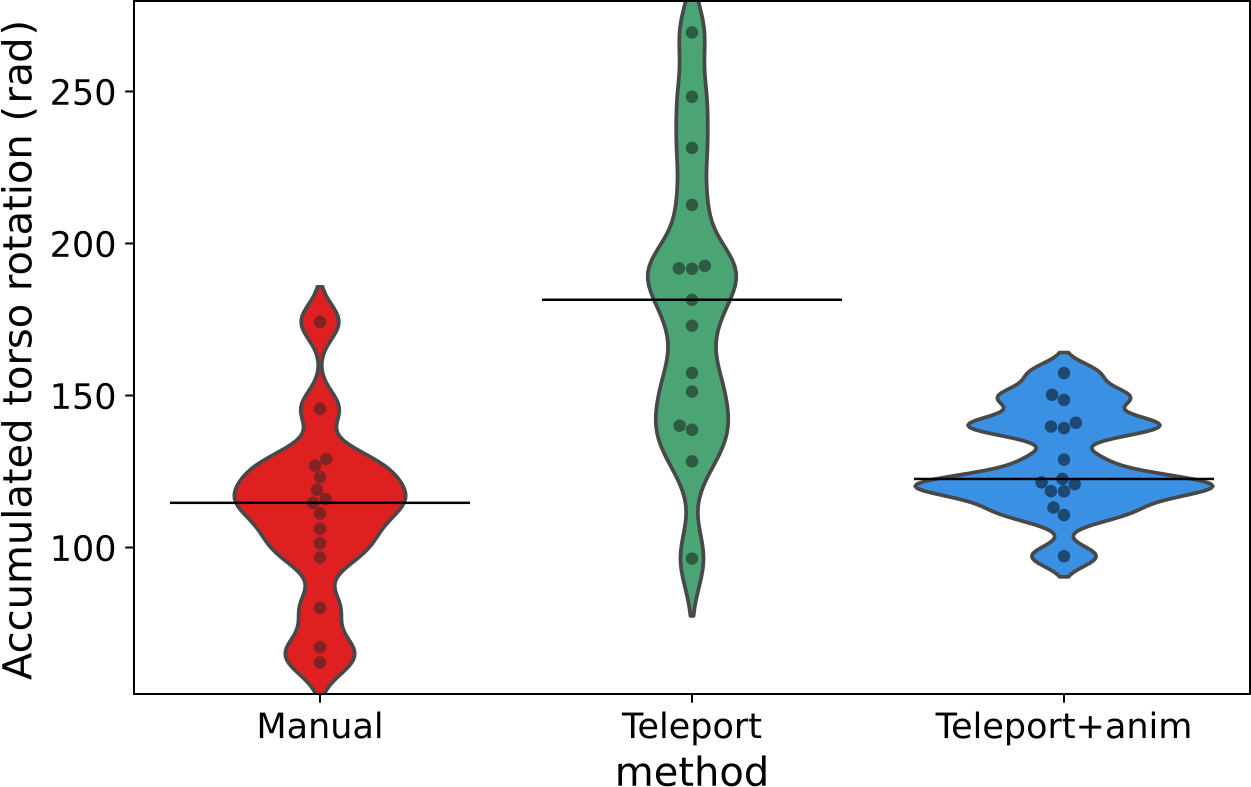}
    \caption{}\label{fig:torr}
    \end{subfigure}
 \caption{Violin plots of the accumulated user's head rotation, torso rotation, and total eye movement. Eye movement was reported in points on the display screen space (range [-1,1] on both axes).}
    \label{}
\end{figure}

For what concerns the number of items reached (e.g. navigation speed), there is a difference among the three conditions (Friedman $chi=10.7826$, $p<0.01$).
Pairwise comparisons using Conover’s all-pairs test with Bonferroni correction show that the most significant difference is between teleport and teleport+animation ($p=0.07$): in the latter condition, the participants ended up reaching fewer items. Yet, this result is most likely influenced by the movement speed that slowed down the whole visit (\autoref{fig:seen}).

The analysis of the number of items correctly located by the users, to measure spatial awareness, (\autoref{fig:located}) showed no significant differences between the three locomotion technique (Friedman $chi=0.800$, $p=0.670$).

Considering the research question $Q3$, there are significant differences in all of the measures among the three conditions (Friedman $chi=19.083$, $p<0.01$, for eye movements; Friedman $chi=39.083$, $p<0.01$ for head angle variations; and, Friedman $chi=9.333$, $p<0.01$ for torso angle variations, in this last case, only the 18 participants with full data has been used in the analysis).
Pairwise comparisons using Conover’s all-pairs test with Bonferroni correction show that the manual technique has significantly fewer eye movements than both teleport ($p<0.01$) and teleport+animation ($p<0.01$), while the two versions of teleport do not have differences (\autoref{fig:eyem}).

The same post-hoc analysis shows that the manual technique has significantly fewer head movements than teleport ($p<0.01$) and teleport+animation fewer than teleport ($p<0.01$). The difference between manual and teleport+animation does not reach significance, but the tendency seems to be that manual is lower (although the standard deviation is pretty high) (\autoref{fig:headr}).

Finally post-hoc analysis for torso angle variations (\autoref{fig:torr}), the only significant difference is that manual is lower than teleport ($p<0.05$) while there is no statistically significant difference for manual vs. teleport+animation and between the two versions of teleport.

\begin{figure}[h]
    \centering
    \includegraphics[width=0.65\linewidth]{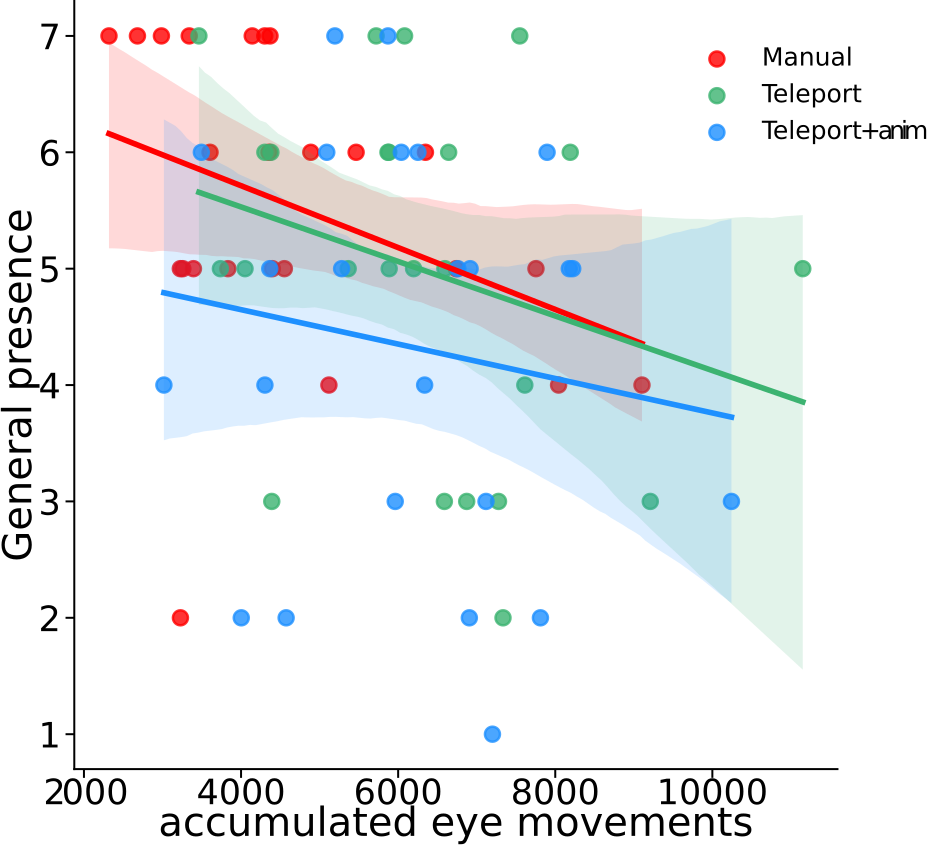}
    \caption{Regression analysis for the three separate methods and the total eye movement on general presence.}
    \label{fig:sp}
\end{figure}

Considering the possibility of explaining (part of) the variation of presence with behavior measures, we fitted a linear mixed model by maximum likelihood for each presence scale (general presence, spatial presence, and involvement) using both eyes movements and head variations (we did not use torso variation because of the missing participants). In order to account for the individual variations, the models included participants as random effects. The only significant model was the one for the general presence scale, and the only significant term was eye movements with a negative score: that is, higher eye movements mean a reduction in spatial presence. A further regression analysis for the three separate conditions (\autoref{fig:sp}) reveals that the effect is significant for the manual condition only and that eye movements explain around 10\% of the variance of the spatial presence in that condition.

A similar approach with perceived difficulty and perceived pleasantness did not produce significant results. We also performed an analysis with all these measures by grouping users by gender; however, no significant difference was found.

\begin{figure}[h!]
    \centering
    \includegraphics[width=0.9\linewidth]{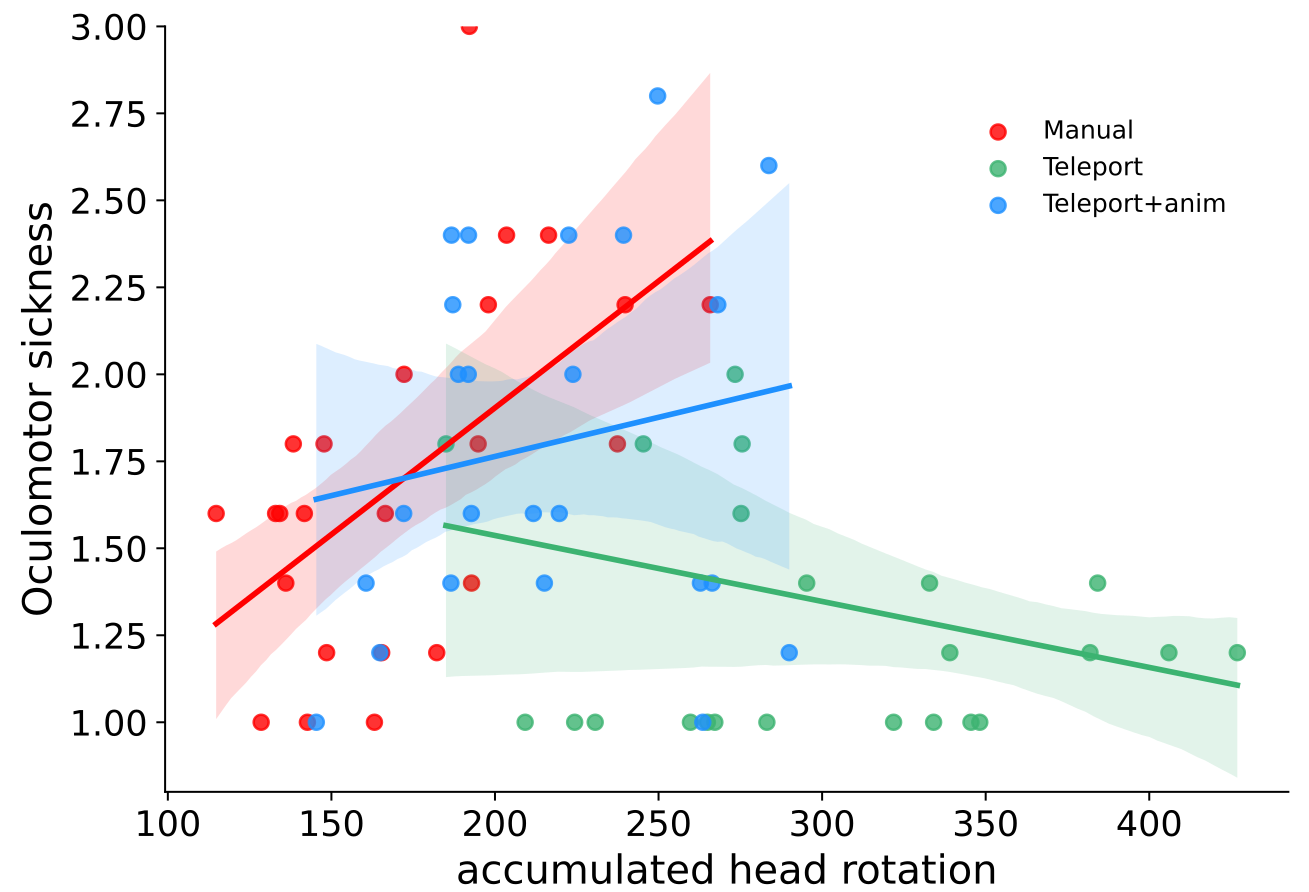}
    \caption{Regression analysis (for the three separate methods) of head rotation vs oculomotor sickness.}
    \label{fig:os}
\end{figure}

Finally, for what concerns behavioural measures and sickness, we performed three separate correlation analyses, one for each technique. The analyses show that only for the manual condition there are significant positive correlations between the head movements and both oculomotor sickness (Pearsons's $r=0.567$, $p<0.05$) (\autoref{fig:os}), and disorientation sickness (Pearsons's $r=0.484$, $p<0.05$); the correlation with nausea sickness is closed to significance (Pearsons's $r=0.375$, $p=0.07$).  

\section{Discussion}

In several studies (see  the discussion of Butussi \& Chittaro \cite{buttussi_locomotion_2021} and Rahimi and colleagues \cite{rahimi2018scene}), teleport was regarded as the best locomotion technique in many respects. Although its value is confirmed by our study too, our results suggest some original insights that might deserve further investigation regarding potential benefits of the manual locomotion technique. In particular, (i) the oculomotor dimensions of sickness seems to be affected more in the manual locomotion than in the others and the movements of the head is a better indicator than the eye movements; (ii) it seems that it might give a higher sense of presence; (iii) the manual technique seems to induce fewer eyes and body movements than the other two.

\subsection{Research question $Q1$}

Our first research question focused on the relation between sickness and susceptibility to sickness, and whether the specific locomotion methods may differently affect this relation.

For what concerns the relation between motion sickness in virtual reality and the (declared) susceptibility, our results do not suggest such a link for either instant teleport or teleport with animation, yet we found a small correlation with the manual technique for what concerns the specific oculomotor dimensions of sickness. This is apparently in contrast to some of the original studies on visual sickness (for example \cite{golding_motion_1998}), but it is worth noting that also a recent literature survey \cite{tian2022review} raises some points of attention. Our results suggest that further studies might need to focus on locomotion techniques as a mediator of the effect. Indeed, the lower accumulated head rotation for the manual method and teleport+animation, suggests that users might be trying to minimize the effects of sickness with techniques involving virtual movement by looking forward and avoiding the creation of optical flow in the foveal region.
Subjects who rotated the head more in manual navigation suffered from increased sickness (\autoref{fig:os}).
We did not find a similar effect for eye movements. Possibly, this is  because, in the point and teleport control, the user spends a lot of time pointing the ray toward a suitable target location and looking at this point.

Regarding the relation between sickness and locomotion technique, our results seem to suggest that the instant teleport technique induces less sickness than the other two techniques. In this respect, it is in line with the previous results  
\cite{langbehn2018evaluation,rahimi2018scene}.

\subsection{Research question $Q2$}
The second research question focused on how the different locomotion techniques impact efficiency, spatial awareness, the perception of presence, difficulty, and pleasantness of the task.

Our results seem to suggest, in line with previous studies 
(see \cite{buttussi_locomotion_2021}),
that subjects perceive the instant teleport as less difficult and more pleasant, at least in comparison with teleport+animation. Nevertheless, the manual technique gives a higher general presence (the sense of being in a place) than teleport+animation, and the tendency suggests that it is higher than teleport (although the difference is not statistically significant). Furthermore, the manual technique is perceived as pleasant as teleport and not more difficult
(in this respect, it might be different from the results of Butussi and Chittaro \cite{buttussi_locomotion_2021}).

We did not find a significant difference in the number of objects correctly located in the post-task questionnaire as this number was in general low with a high variance. However, the fact that the average count for the animated teleport is higher for the animated teleport (with a non-negligible difference) may suggest further investigation into spatial awareness as a function of visual feedback.

It is worth noting that, contrary to other studies (see \cite{rahimi2018scene}), we did not find differences in the spatial dimension of presence among the methods. In this respect, we can say that for the users, there's no trade-off between reduced sickness by using teleport and better mapping of the environment by using techniques featuring continuous viewpoint animation.

Our results may suggest that manual technique may be a valid alternative to teleporting in some cases. This was not found by Butussi and colleagus \cite{buttussi_locomotion_2021} where the authors did not measure differences in presence across the tested methods and possibly related to the different tasks proposed. 

\subsection{Research question $Q3$}

The third research questions was aimed at understanding whether the eyes and body movements are differed  among the conditions, and whether those changes can explain presence, difficulty, or pleasantness.

Indeed, the manual technique seems to induce fewer eyes and body movements than the other two. Given the well-established relation between body and eye movements to sickness \cite{stoffregen_instability_2002,stoffregen_instability_2002,munafo_virtual_2017}, these results seem to suggest that, at least in some situations, the manual technique might be preferred. Although we could not find any relation between body movements and sickness,  there was a small but significant negative correlation between the variation of the head angle and all three measures of sickness: yet, in the manual condition only. 

During the trials, we observed that many participants, while moving with either teleport method, preferred to only rotate their torso and head to aim the selection ray rather than turning on their feet to face the intended direction. This fact suggests that some other effect, such as the fear of moving in the real world while unable to see it, can influence the interaction and therefore needs further investigation. Yet, as a matter of speculation, the increased head and body movements for teleporting may be due to the necessity of recovering orientation after the viewpoint displacement. This recovery seems effective in keeping spatial awareness while not causing sickness, but we could not bring evidence in support, as the view is rotated but does not translate. On the other hand, the positive correlation between head rotation and sickness in the manual condition reveals that the sickness effects may be due to lateral optical flow in the foveal region while translating forward and suggests that potential solutions to reduce it can rely on specific training or visual feedback helping the user to avoid head rotation when moving. Rotations while translating forward are allowed in animated teleport as well, and the fact that it was not related to sickness in \cite{bhandari2018teleportation} is possibly due to the lack of scene texture creating the optical flow.

Although the evidence is still preliminary, an original result of our study is that, for the manual technique only, the amount of eye movements (but not of the movements of the head and body) seems correlated with decreased general presence. Nevertheless, for the other dimensions of experience, such as difficulty and pleasantness, we did not find any correlation, and this suggests that further studies are still needed. 

Considering the goal of better understanding the relationship between vection and sickness \cite{keshavarz2015vection}, our results are inconclusive, and it is probably necessary to capture further information during the tasks, like camera motion, magnitude and localization of the optical flow on the image plane to have a better understanding of the phenomenon. In future studies, we also plan to extract additional information from the eye-tracking records (e.g., saccades) and estimate postural instability with external tracking.

\color{black}

Finally, it is worth noting that, contrary to other studies (for example, \cite{munafo_virtual_2017}), in our case, there is no difference between males and females in either susceptibility to sickness or the actual sickness experienced in the three conditions. We collected this data in order to prevent an impact of this variable on the analysis.

\section{Conclusion}
In this work, we presented the results of a within-subject study investigating the effects of three locomotion techniques on the quality of an immersive VR experience, including presence, spatial awareness, and sickness. 
We also tried to find correlations between those measures and the behavior in terms of eyes, head, and body movements.

Our study confirms some results already discussed in the literature regarding the reduced invasiveness and the high usability of instant teleport \cite{buttussi_locomotion_2021} and some of the issues related to teleporting with animation \cite{rahimi2018scene}.  Contrary to the results of Bowman and colleagues \cite{bowman1997travel} 
and Rahimi et al. \cite{rahimi2018scene}, 
and in accordance with Bhandari et al. \cite{bhandari2018teleportation}, and Langbehn et al. \cite{langbehn2018evaluation}, we did not find that teleporting increase disorientation in terms of spatial awareness or reduced spatial presence. 

Also, contrary to what seems to be suggested by Langbehn et al. \cite{langbehn2018evaluation} our study suggests that the manual locomotion technique might have benefits of both instant and animated teleporting with less problematic aspects.

An original result worth further investigation is that, for what concerns the user experience (i.e. general presence, at least pleasantness, and perceived difficulty of the task), both the manual and the instant teleport techniques score better than teleport with animation and, although there are no significant differences among them, the tendency appears to favor the manual technique. 
\color{black}

We plan to further investigate these effects in future works and to further exploit the additional data acquired from subjects' behavior and physical response.
 
\section*{Acknowledgments}
This study was partially partially carried out within the PNRR research activities of the consortium iNEST (Interconnected North-Est Innovation Ecosystem) funded by the European Union Next-GenerationEU (Piano Nazionale di Ripresa e Resilienza (PNRR) – Missione 4 Componente 2, Investimento 1.5 – D.D.~1058 23/06/2022, ECS00000043).

%
%
%
%
\bibliographystyle{splncs04}
\bibliography{againteleporting.bib}

\end{document}